\documentclass[12pt]{article}

\catcode`\@=11
\global\arraycolsep=2pt
\usepackage{amsbsy,amssymb,latexsym,amsfonts,amsmath}
\usepackage{graphicx,color}

\begin{document}
\begin{flushright}
\parbox{4.2cm}
{RUP-20-01}
\end{flushright}

\vspace*{0.7cm}

\begin{center}
{ \Large On Duality in $\mathcal{N}=2$ supersymmetric Liouville Theory}
\vspace*{1.5cm}\\
{Yu Nakayama}
\end{center}
\vspace*{1.0cm}
\begin{center}

Department of Physics, Rikkyo University, Toshima, Tokyo 171-8501, Japan

\vspace{3.8cm}
\end{center}

\begin{abstract}
Similarly to the bosonic Liouville theory, the $\mathcal{N}=2$ supersymmetric Liouville theory  was conjectured to be equipped with the duality that exchanges the superpotential and the K\"ahler potential. The conjectured duality, however, seems to suffer from a mismatch of the preserved symmetries. More than fifteen years ago, when I was a student, my supervisor Tohru Eguchi gave a beautiful resolution of the puzzle  when the supersymmetry is enhanced to $\mathcal{N}=4$ based on his insight into the underlying geometric structure of the $A_1$ singularity. I will review his unpublished but insightful idea and present our attempts to extend it to more general cases.
\end{abstract}

\thispagestyle{empty} 

\setcounter{page}{0}

\newpage


%-----------------------------------------

%\pacs{}
%\keywords{Suggested keywords}%Use shokeys class option if keyword
%\maketitle

%%%%%%%%%%%%%%%%%%%%%%%%%%%%%%%%%%%%%%%%%%%%%%%%%%%%
\section{Introduction}

My master's thesis was about the Liouville theory \cite{Nakayama:2004vk}.\footnote{In Japan, every grad student must write a master's thesis. In our community, most of them are reviews of contemporary topics.} In my thesis defense, I chose to discuss the dualities in the Liouville theory. In particular, I tried to elucidate the $\mathcal{N}=2$ supersymmetric version of the Liouville duality. This was a fascinating topic to me because it claims that the theory is invariant under the exchange of the superpotential and the  K\"ahler potential. We were taught, at least in four dimensions, that we can non-perturbatively study the superpotential but we cannot tame the K\"ahler potential.  I was dreaming that studying this duality further might lead to a revolution in our understanding of supersymmetric field theories.

At the time, I was thinking thesis defense is a fun event just to present what I have learned and what I find interesting. But then, a horrible moment came when Yanagida-san raised his hand and said that this duality cannot be true. He continued that he can immediately see that the shift symmetry that is apparent in one frame of the claimed duality is manifestly broken in the other frame of the duality. He instinctively knew this because, back then, the shift symmetry in the K\"ahler potential was believed to be a crucial ingredient to support the inflation in supergravity (since otherwise, the potential is exponentially growing at the Planck scale) and he was vigorously studying this mechanism. I should have known it, but it was too late. I felt desperate.

Well, after all, I barely passed the defense: I had been aware of this difficulty myself before, so I tried to explain it by using more sophisticated mathematics (i.e. a mirror symmetry of Hori and Kapustin here). 
But nobody in the room, including me, was satisfied. It was apparent that something is wrong with this ``duality". I don't remember if there was a comment by Eguchi-san on the spot. He was my supervisor and was a chair of the thesis committee.

Given this experience, I stopped thinking about this (unsatisfactory) ``duality" in the $\mathcal{N}=2$ supersymmetric Liouville theory. My dream of taming the K\"ahler potential had faded away subsequently and I was looking for other subjects to study.
Several months later, however, I got a sudden email from Eguchi-san with a note, which said that he resolved this puzzle. As a student of his, I was presumably supposed to work with him on it.  This article aims to review this unpublished note by Eguchi-san on June 9th, 2004. Looking back now, I think I should have pursued this question further with him. I sincerely regret that we have lost the opportunity forever.

\section{Duality in bosonic Liouville theory}
The quantum Liouville theory conventionally described by the classical action
\begin{align}
S_b = \frac{1}{4\pi} \int d^2x \left (4\partial \phi \bar{\partial}\phi - Q R \phi + 4\pi \mu e^{-2b\phi} \right)
\end{align}
is an exactly solved conformal field theory in two-dimensions.\footnote{For a historical reason, we use the so-called $\alpha'=1$ convention in the bosonic Liouville theory. We will switch to the $\alpha'=2$ convention from section 3. To take into account Eguchi-san's preference, the sign of the Liouville exponent in this  article is chosen to be opposite to the one used in my master's thesis \cite{Nakayama:2004vk}.} The Liouville theory has one parameter $b$, which is related to the background charge $Q= b+b^{-1}$ appearing in the dilaton coupling (i.e. so-called the Fradkin-Tseytlin term) and determines the central charge $c = 1 + 6(b+b^{-1})^2$ with the holomorphic energy-momentum tensor
\begin{align}
T = -(\partial\phi)^2 - Q\partial^2 \phi \ .
\end{align} 
The classical limit is given by $b\to 0$, in which the path integral based on the above action is reliable,\footnote{To see this, we may introduce the classical Liouville field $\varphi = b\phi$ and regard $b^2$ as $\hbar$,} and the classical saddle point can be studied by solving the classical Liouville equation which Joseph Liouville introduced in the 19th century.

As the expression of the central charge may suggest, the quantum  Liouville theory shows a ``duality" under the exchange of $b$ and $b^{-1}$. Indeed, if we examine the exact expressions for two-point functions and three-point functions \cite{Dorn:1994xn}\cite{Zamolodchikov:1995aa}\cite{Teschner:1995yf}, we realize that they are all invariant under the exchange of $b$ and $b^{-1}$ if we further replace the Liouville cosmological constant $\mu$ with the dual one $\tilde{\mu}$ related by 
\begin{align}
\tilde{\mu} = \frac{(\pi\mu \gamma(b^2))^{b^{-2}}}{\pi \gamma(b^{-2})} \ ,
\end{align}
where $\gamma(x) = \Gamma(x)/\Gamma(1-x)$.
This means that we may well quantize the theory based on the ``dual Liouville action"
\begin{align}
S_{b^{-1}} = \frac{1}{4\pi}  \int d^2x \left (4\partial \phi \bar{\partial}\phi - Q R \phi + 4\pi \tilde{\mu} e^{-2b^{-1}\phi} \right)
\end{align}
instead as a starting point. In the dual picture, the classical limit is $b \to  \infty$. At $b=1$ (with $c=25$), we have a self-duality and this fact plays an important role to understand the $c=1$ string theory.

This duality structure reminds us of the dual screening charge of the Coulomb gas formalism except that the ``screening momentum" is imaginary here (i.e. $e^{ip X}$ v.s. $e^{\alpha \phi}$). During the development of the exact solutions of the quantum Liouville theory \cite{Dorn:1994xn}\cite{Zamolodchikov:1995aa}\cite{Teschner:1995yf}, it was often assumed that the both Liouville potential and the dual Liouillve potential appear in the classical action and both can be used freely to do perturbative computations (as is the case in the Coulomb gas formalism). In other words, we may study the ``perturbative" correlation functions as if we have the combined action:
\begin{align}
S_{b+b^{-1}} = \frac{1}{4\pi}  \int d^2x \left (4\partial \phi \bar{\partial}\phi - Q R \phi  + 4\pi\mu e^{2b\phi} + 4\pi \tilde{\mu} e^{-2b^{-1}\phi} \right) \ . 
\end{align}

The available exact solutions are consistent with this picture (partly because they are originally obtained under this duality assumption). For example, the poles in the two-point and three-point functions are located when the background charges are screened by the ``perturbation" by the Liouville potential {\it and} the dual Liouville potential {\it simultaneously}. However, it remains an open question to understand what we really mean by ``adding the dual Liouville potential" to the classical action. We stress that once the consistency and possible uniqueness of the Liouville correlation functions are verified in the sense of the conformal bootstrap, we do not need any ``perturbative" picture based on the action principle nor the path integral formalism.

It is therefore understandable that some of our colleagues do not like the idea of adding the dual Liouville potential \cite{Ribault:2014hia}\cite{Kupiainen:2018snr}: they prefer that the quantization should be done in one or the other duality frame. This is deeply related to a rather philosophical question of what we actually mean by the path integral quantization based on a classical action when the exact result is available. We will not go into the discussions further here, but we have something  to say about a related issue in the $\mathcal{N}=2$ supersymmetric Liouville theory.

\section{Duality in $\mathcal{N}=2$ supersymmetric Liouville theory at $Q=1$}

\subsection{Proposed duality and a puzzle}
Let us consider the $\mathcal{N}=2$ supersymmetric Liouville theory \cite{Ivanov:1983wp}. The free part of the action is given by
\begin{align}
 S = \frac{1}{2\pi}\int d^2x \left( \partial \phi \bar{\partial}\phi +   \partial Y \bar{\partial}Y  - \frac{1}{4} QR\phi + {\Psi}^+ \bar{\partial} \Psi^- + \bar{\Psi}^+ \partial \bar{\Psi}^- \right) \ .
\end{align}
Here $\phi$ is the Liouville field with a background charge $Q$, and $Y$ is a compactified boson whose radius will be specified below. The superpartners $\Psi^{\pm}$ and $\bar{\Psi}^{\pm}$ are (left-moving and right-moving) Dirac fermions. The central charge is given by
\begin{align}
c = 3 \hat{c} = 3(1+Q^2) \ .
\end{align}

The theory admits the $\mathcal{N}=2$ superconformal symmetry generated by the holomorphic current 
\begin{align}
T & = -\frac{1}{2} (\partial Y)^2 -\frac{1}{2} (\partial \phi)^2 - \frac{Q}{2} \partial^2 \phi - \frac{1}{2} \left(\Psi^+ \partial \Psi^- -\partial \Psi^+ \Psi^- \right) \cr
G^{\pm} &= -\frac{1}{\sqrt{2}} \Psi^{\pm}(i\partial Y \pm \partial \phi) \mp \frac{Q}{\sqrt{2}} \partial \Psi^{\pm} \cr
J &= \Psi^+ \Psi^- - Q i \partial Y \ . 
\end{align}
and their anti-holomorphic partners.

We have two types of Liouville potentials: one is given by the superpotential
\begin{align}
S_{+}& = \int d^2 x d^2 \theta Q^2 e^{-\frac{1}{Q} \Phi} = \int d^2x  \Psi^{-} \bar{\Psi}^{+} e^{-\frac{1}{Q}(\phi+iY)} \cr
S_{-} &=  \int d^2 x d^2 \bar{\theta} Q^2 e^{-\frac{1}{Q} \bar{\Phi}} = \int d^2x  \Psi^+ \bar{\Psi}^{-} e^{-\frac{1}{Q}(\phi-iY)}
\end{align}
and the other is given by the K\"ahler potential:
\begin{align}
S_{3} &= \int d^2 x d^4 \theta Q^{-2} e^{-\frac{Q}{2}(\Phi + \bar{\Phi})} \cr
&= \int d^2x \left(\partial \phi - i\partial Y -Q\Psi^+\Psi^-\right)\left(\bar{\partial}{\phi} + i \bar{\partial}Y + Q \bar{\Psi}^+ \bar{\Psi}^- \right) e^{-Q\phi} \ .
\end{align}
Here we have used the superfield formalism $\Phi = \phi + i Y + \cdots$ to make the $\mathcal{N}=2$ supersymmetry manifest.\footnote{Note that the contact terms from the auxiliary fields are omitted as usual in the supersymmetric Liouville literature. See e.g. \cite{Hosomichi:2004ph}.}

For later purposes, let us discuss the preserved symmetries under these Liouville interactions. For generic $Q$ without the Liouville interactions, in addition to the $\mathcal{N}=2$ superconformal symmetry, we have the symmetry associated with the shift of $Y$ and the winding of $Y$ (whose currents are generated by $\partial Y$ and $\bar{\partial} Y$). The both $S_{\pm}$ and $S_3$ preserve the $U(1)_R$ symmetry, but the former break the shift of $Y$, while the latter preserves all the above-mentioned symmetries. 

The $\mathcal{N}=2$ supersymmetric Liouville theory is exactly solved \cite{Ahn:2002sx}\cite{Baseilhac:1998eq}\cite{Hosomichi:2004ph}. If we examine the correlation functions of the $\mathcal{N}=2$ supersymmetric Liouville theory, we realize that they have poles if the background charges are screened both by $e^{-Q\phi}$ and $e^{-Q^{-1}\phi}$. This may suggest a duality between the superpotential and the K\"ahler potential. The nature of the duality, however,  seems more complicated than the bosonic case discussed in the previous section. In particular, it is not immediately obvious which of the following is the correct interpretation of the duality: (1) it is the duality between the theory with the superpotential but without the K\"ahler potential and the theory with the K\"ahler potential but without the superpotential, or (2) we should add the both at the same time to define the $\mathcal{N}=2$ supersymmetric Liouville theory.

Indeed, we can immediately argue against the literal sense of the duality between the superpotential and the K\"ahler potential because the former breaks the shift symmetry of $Y$ while the latter preserves it. To avoid this issue, in almost all the literature studying the duality of the $\mathcal{N}=2$ supersymmetric Liouville theory, they only studied the so-called zero-charge sector (with respect to the shift of $Y$) with (partial) success  \cite{Ahn:2002sx}. We could have taken the view that the K\"ahler potential must have been added (in addition to the superpotential) as a definition of the $\mathcal{N}=2$ supersymmetric Liouville theory, but this also raises the question what we really mean by the $\mathcal{N}=2$ supersymmetric Liouville theory as in the bosonic Liouville theory discussed in the previous section.

As we have mentioned in the introduction, this duality is related to a mirror symmetry by the work of Hori and Kapustin \cite{Hori:2001ax}. The duality proposed by Hori and Kapustin claims the following: suppose we compactify $Y$ to its minimal radius compatible with the Liouville superpotential (i.e. $Y \sim Y + 2\pi Q$), and it is dual to the $SL(2,\mathbb{R})/U(1)$ Kazama-Suzuki supercoset model \cite{Kazama:1988uz} with the level $k=\frac{2}{Q^2}$. It is a duality in the sense of the mirror symmetry because the $U(1)_R$ symmetry is left-right flipped. In the string theory, the A-model (or B-model) on the $\mathcal{N}=2$ supersymmetric Liouville theory is the same as the B-model (or A-model) on the $SL(2,\mathbb{R})/U(1)$ supercoset model. 

We will not go into the details of this duality or its derivation here, but let us discuss the consequence and its relation to the duality of the $\mathcal{N}=2$ supersymmetric Liouville theory. For large $k$, the Kazama-Suzuki supercoset model has a sigma model (geometrical) description. The classical target space is given by a cigar
\begin{align}
ds^2 =k(  d\rho^2 + \tanh^2(\rho)d\theta^2) \ .
\end{align}
with the background dilaton
\begin{align}
\Phi = -2\log \cosh \rho  \ .
\end{align}
Here periodicity of $\theta$ is taken to be $2\pi$.

It is important to address the fate of the shift symmetry broken by the Liouville superpotential in terms of the dual picture. In the mirror description, the shift symmetry becomes a winding symmetry along the $\theta$ direction. However, the winding in $\theta$ direction is not conserved because the cigar has a trivial first homotopy class. One can unwind the string at the tip of the cigar. Thus, the winding number conservation is ``non-perturbatively" broken, which is the dual statement of the non-conservation of the shift of $Y$ in the $\mathcal{N}=2$ supersymmetric Liouville theory. Here the $\theta$ coordinate is identified as the mirror-dual of the $Y$ coordinate.

Now, to relate this mirror symmetry with the $\mathcal{N}=2$ supersymmetric Lioville duality, let us perform the formal application of Buscher's T-duality rule to the periodic $\theta$ direction. Then the resultant geometry is given by
\begin{align}
ds^2 = k \left( d\rho^2 + \tanh^{-2}(\rho)d\tilde{\theta}^2\right) \ , \label{dualm} 
\end{align}
where $\tilde{\theta}$ is the dual coordinate with the periodicity $2\pi/k^2$.
If we expand the geometry around $\rho \to \infty$, we realize that the leading deviation from the flat cylinder with a linear dilaton is nothing but the $\mathcal{N}=2$ Liouville K\"ahler potential. 

This observation may explain the origin of the duality in the  $\mathcal{N}=2$ supersymmetric Liouville theory. However, the question about the symmetry breaking pattern remains. In the $SL(2,\mathbb{R})/U(1)$ supercoset model, the winding symmetry is broken by the non-perturbative effect. How is it realized in the T-dual geometry? Naively, the metric \eqref{dualm} has a $U(1)$ isometry, preserving the momentum along $\tilde{\theta}$.  In the T-dual geometry, $\rho=0$ becomes a singularity and the classical prediction will be lost and this is probably how the seemingly conserved shift symmetry in $\tilde{\theta}$ direction, which is supposed to be identified with $Y$ (after rescaling), is broken. But how? Even if true, the more urgent question is in the duality of the  $\mathcal{N}=2$ supersymmetric Liouville theory, we have added the only leading term of the geometric deformation and we do not see a hint of the singularity at all. Is the proposed duality just an approximation of the more fundamental duality?

\subsection{A resolution of the puzzle at $Q=1$}
A beautiful resolution of the puzzle at $Q=1$ was proposed by Eguchi-san in 2004, which I will review now.\footnote{The partial result was presented by Eguchi-san in \cite{Eguchi:2004kx}. After the first version of this note appeared on arXiv, I realized that \cite{Murthy:2003es} addressed the same problem and the resolution from the viewpoint of the system of two NS5-branes, which is T-dual to the Eguchi-Hanson space we will discuss below. I would like to thank S.~Murthy for the correspondence.} His idea was based on the geometric intuition of the  hyper K\"ahler structure of the Eguchi-Hanson space \cite{Eguchi:1978xp}. The Eguchi-Hanson space describes the $A_1$ singularity of a  complex two-dimensional surface. It admits a Ricci-flat metric with an $SU(2)$ holonomy (i.e. two-dimensional Calabi-Yau space). The crucial feature we would like to employ here is that it is a hyper-K\"ahler manifold: the nowhere-vanishing holomoprhic  two-form $\Omega$, the anti-holomorphic two-form $\bar{\Omega}$ and the K\"ahler two-form $K$ transform as a triplet under the hyper-K\"aler rotation. Accordingly, if we study a (type II) superstring theory on the Eguchi-Hanson space, it must have $\mathcal{N}=4$ (rather than $\mathcal{N}=2$) worldsheet  superconformal symmetry. 

The $\mathcal{N}=2$ supersymmetric Liouville theory with $Q=1$ has the central charge $c=6$ and it is an appropriate worldsheet theory that can be used in the string compactification. The study  of \cite{Giveon:1999px}\cite{Giveon:1999tq}\cite{Eguchi:2000tc}\cite{Eguchi:2004ik} suggests that the $\mathcal{N}=2$ supersymmetric Liouville theory as a worldsheet string theory describes the string theory on the Eguchi-Hanson space. As we have just mentioned, the worldsheet string theory must have the enhanced $\mathcal{N}=4$ superconformal symmetry. How are they realized? This becomes the starting point of our discussions.

Consider the $\mathcal{N}=2$ supersymmetric Liouville theory at $Q=1$ with the $\mathcal{N}=2$ superconformal current given by
\begin{align}
T & = -\frac{1}{2} (\partial Y)^2 -\frac{1}{2} (\partial \phi)^2 - \frac{1}{2} \partial^2 \phi - \frac{1}{2} \left(\Psi^+ \partial \Psi^- -\partial \Psi^+ \Psi^- \right) \cr
G^{\pm} &= -\frac{1}{\sqrt{2}} \Psi^{\pm}(i\partial Y \pm \partial \phi) \mp \frac{1}{\sqrt{2}} \partial \Psi^{\pm} \cr
J &= \Psi^+ \Psi^- -  i \partial Y \ . 
\end{align}
To see the enhancement to the $\mathcal{N}=4$ superconformal symmetry, it is convenient to bosonize the Dirac fermions:\footnote{I would rather fermionize $Y$, but I will stick to Eguchi-san's note here.}
\begin{align}
\Psi^+(z) \Psi^-(z) &= i\partial H(z) \cr
\bar{\Psi}^+(\bar{z}) \bar{\Psi}^-(\bar{z}) &= i\bar{\partial} H(\bar{z}) \cr 
\Psi^+(z) &= e^{iH(z)} \cr
\Psi^-(z) &= e^{-iH(z)} \ .
\end{align}
Note that if we compactify $Y$ at the minimum radius at $Q=1$, which we do here, then $H$ and $Y$ have the same radius. In other words, $Y$ is at the free fermion point (without the Liouville potential). This is a good sign because the $\mathcal{N}=4$ superconformal symmetry will rotate them eventually. 

Now one can construct the conserved $SU(2)$ R-current:
\begin{align}
J^+_{SU(2)} &= e^{iH-iY} \cr
J^-_{SU(2)} & = e^{-iH+iY} \cr
J^3_{SU(2)} & = \frac{1}{2}(i\partial H - i \partial Y) \ 
\end{align}
as well as the $\mathcal{N}=4$ supercurrent:
\begin{align}
G^{+,+} &= G^+ = -\frac{1}{\sqrt{2}} \Psi^{\pm}(i\partial Y \pm \partial \phi) \mp \frac{Q}{\sqrt{2}} \partial \Psi^{\pm} \cr
G^{-,-} &= G^- = -\frac{1}{\sqrt{2}} \Psi^{\pm}(i\partial Y \pm \partial \phi) \mp \frac{Q}{\sqrt{2}} \partial \Psi^{\pm} \cr
G^{+,-} &= J^+_{SU(2)}\cdot G^{-,-} = \frac{1}{\sqrt{2}} e^{-iY}(\partial\phi - i\partial H) + \frac{1}{\sqrt{2}} \partial e^{-iY} \cr
G^{-,+} &= J^{-}_{SU(2)} \cdot G^{+,+} = -\frac{1}{\sqrt{2}} e^{iY} (\partial \phi + i \partial H) - \frac{1}{\sqrt{2}} \partial e^{iY} \ . 
\end{align}
The $\cdot$ operation means to take the residue of $1/z$ in the OPE. 

With the bosonization at $Q=1$, the Liouville interactions become
\begin{align}
S_+ &= \int d^2x e^{-\phi - iY - iH} \cr
S_3 &= \int d^2x \left(\partial\phi - i \partial Y - i\partial H \right)\left(\bar{\partial} \phi + i \bar{\partial}Y + i \bar{\partial} H \right) e^{-\phi} \cr
S_- &= \int d^2x e^{-\phi + i Y + iH} \ . 
\end{align}
We emphasize that they all share the same Liouville exponent of $e^{-\phi}$ at $Q=1$. All of them are compatible with the above constructed $\mathcal{N}=4$ superconformal symmetry.

Here is a crucial observation. The $\mathcal{N}=4$ superconformal algebra admits another $SU(2)_{\text{outer}} = SU(2)'$ algebra acting as an outer automorphism. Explicitly, they are generated by
\begin{align}
J^+_{SU(2)'} &= e^{iY +iH} \cr
J^3_{SU(2)'} &= \frac{1}{2}(i\partial H + i \partial Y) \cr
J^-_{SU(2)'} &= e^{-iY - iH} \ 
\end{align}
without the Liouville interactions. Under this $SU(2)_{\text{outer}}$, the Liouville interactions $S_{\pm}$, $S_3$ are not invariant but form a triplet. 

The action of the $SU(2)'$ on $\mathcal{N}=4$ supercharges as an outer automorphism are
\begin{align}
G^{+,-} &= J^{-}_{SU(2)'} \cdot G^{+,+} \cr
G^{-,+} &= J^{+}_{SU(2)'} \cdot G^{-,-} 
\end{align}
so that $SU(2)$ act on the left index of $G^{a,b}$ while $SU(2)'$ act on the right index. Thus, without the Liouville interaction, the theory has $SU(2) \times SU(2)'$ algebra at $Q=1$ (together with the anti-holomorphic copy) but the $SU(2)'$ symmetry is broken by the presence of the Liouillve interaction. The $SU(2)'$ action rotates $S_{\pm}$ into $S_3$ and vice versa. This means that at $Q=1$ the $\mathcal{N}=2$ supersymmetric Liouville theory with the superpotential deformation is completely equivalent to the $\mathcal{N}=2$ supersymmetric Liouville theory with the K\"ahler potential deformation because they are rotated by the $SU(2)'$, providing a proof of the conjectured duality.

A couple of comments are in order. The most urgent concern about the conjectured duality was the symmetry breaking pattern. The above explanation resolves the issue. In each deformations, $SU(2)'$ rather than $U(1) \times U(1)$ is broken down to $U(1)$.\footnote{To be more precise with the anti-holomorphic sector, we have $SU(2)'_L \times SU(2)'_R$ broken down to $SU(2)$ (which contains the winding number $U(1)_w$).} In particular, the K\"ahler deformation does break the symmetry that rotates $H$ and $Y$, and we do not have the mismatch of the symmetry breaking pattern albeit they are not manifest in the language of $\mathcal{N}=2$ supersymmetry.

The second point is that the duality is a manifestation of the hyper K\"aher structure of the Eguchi-Hanson space. When we resolve the $A_1$ singularity, there are apparently two different ways, the one by the K\"ahler deformation and the other by the complex structure deformation. They each other correspond to adding the Liouville K\"ahler potential or to adding the Liouville superpotential. However, we know that the Eguchi-Hanson space is hyper K\"ahler and they are physically equivalent. In other words, the duality in the $\mathcal{N}=2$ supersymmetric Liouville theory (at $Q=1$) is nothing but the hyper K\"ahler rotation of the Eguchi-Hanson space. 

Finally, this picture provides us with a fresh viewpoint on the nature of the duality. Namely, we could have a debate whether we should add the three interactions simultaneously or we should regard them as two different theories related by the duality. The discussions here suggest it is irrelevant. We could have added the K\"ahler potential deformation and the superpotential deformation simultaneously if we want, but this is simply choosing a particular direction of the  $SU(2)'$ which will be broken, so they are completely equivalent to choosing either the K\"ahler potential deformation alone or the superpotential deformation alone. 

\section{Generalization to $Q\neq 1$}
The beautiful story ends here, and homework was left. Eguchi-san was very optimistic about the generalization to the $Q\neq 1$ case. As we will see, it is not that immediate and it should not be so. After all, we will not be blessed by the beauty of $\mathcal{N}=4$ superconformal symmetry and the hyper K\"ahler rotation. We, nevertheless, attempt to generalize the idea so that we can learn something about the nature of duality in $\mathcal{N}=2$ supersymmetric Liouville theory.\footnote{An alternative way is to keep the $\mathcal{N}=4$ superconformal symmetry by adding the $SU(2)$ sector. This will lead to the more general $\mathcal{N}=4$ supersymmetric Liouville theory \cite{Matsuda:1996bb}\cite{Eguchi:2016cyh}, and then there naturally exists a duality corresponding to the hyper K\"ahler rotation of higher $A_n$ singularities (see e.g. \cite{Murthy:2003es} for the related discussions).  We will, however, focus on the $\mathcal{N}=2$ case here.} 

Let us try to deform the $\mathcal{N}=4$ supersymmetric Liouville theory to $\mathcal{N}=2$ supersymmetric Liouville theories by changing the background charge $Q$ from the $\mathcal{N}=4$ value of $Q=1$. At the free theory level, we can do this by adding the background charge changing operator
\begin{align}
\frac{1}{2\pi}\int d^2x \left( \alpha \partial \phi \bar{\partial}\phi - \beta R \phi + \gamma {\partial}Y \bar{\partial}Y \right) 
\end{align}
to the action. In order to see the effect of the deformations, we have to normalize the kinetic term by setting $\phi \to \frac{1}{\sqrt{1+\alpha}} \phi$. Then the central charge is modified to 
\begin{align}
c = 3\left(1+\frac{(1+4\beta)^2}{1+\alpha} \right) \ .
\end{align}
Note that at the level of the free theory only a particular combination of $\alpha$ and $\beta$ is physically meaningful. In other words, the above deformation contains a redundant operator. Note that the effect of $\gamma$ is to change the compactification radius of $Y$ (by a factor of $1/\sqrt{1+\gamma}$), so we should think it is physical rather than redundant. 

With the Liouville interaction, the discussions become more complicated. For our purpose, let us choose the superpotential deformations (i.e. $S_{\pm}$) as our starting point of the $\mathcal{N}=2$ supersymmetric Liouville theory (at $Q=1$). Then the background charge changing deformation we have to add to obtain the $\mathcal{N}=2$ supersymmetric Liouville theory with a generic $Q$ is
\begin{align}
\frac{1}{2\pi} \int d^2x (Q^2-1)\left (\partial \phi \bar{\partial}\phi + \partial Y \bar{\partial}Y - \frac{1}{4} R \phi \right) \ . \label{inter}
\end{align}
Note that the parameters $\alpha$, $\beta$ as well as $\gamma$ are fixed by demanding the superconformal invariance of the Liouville superpotential $W = e^{-\Phi} \to e^{-Q^{-1}\Phi}$ after making the Liouville fields canonically normalized.

What will be the corresponding deformations in the dual frame?
The naive idea is to first rotate the Liouville interaction at $Q=1$ to any directions that we like (say the K\"ahler potential deformation by $S_3$) and add the interaction \eqref{inter} to make $Q$ arbitrary, but this does not work for several (obvious) reasons. If it were successful, we would get back to the original question of the mismatch of the symmetry breaking pattern: the superpotential breaks the shift symmetry of $Y$ but the K\"ahler potential does not break it. 

A closer look reveals another difficulty: with the K\"ahler potential deformation, the background charge changing deformation \eqref{inter} does not preserve the superconformal symmetry because in the canonical normalization of Liouville fields, the Liouville exponent in the K\"ahler potential deformation should be $e^{-Q\phi}$ rather than $e^{-Q^{-1} \phi}$ as the application of \eqref{inter} would do (after making kinetic terms canonically normalized). In other words, the redundant deformations we had without the superpotential deformation or K\"ahler potential deformation are not compatible with each other.

We have to overcome these points. As for the first point, since we rotate the Liouillve interactions by the $SU(2)'$, we also have to rotate the background charge deformation  by the same $SU(2)'$.  By construction, this should resolve the mystery of the symmetry breaking pattern although the resultant ``duality" may be different from the original one. While the final result will be necessarily different from the duality of the $\mathcal{N}=2$ supersymmetric Liouville theory originally conjectured for $Q\neq 1$, it is worth pursuing further.\footnote{In particular, the preserved $\mathcal{N}=2$ superconformal symmetry is different.} For definiteness we will try to completely rotate the Liouville interaction so that we would end up with $S_3$. 

Still, we have to face the second problem. 
Let us first note that $(-1+Q^2)({\partial}\phi \bar{\partial} \phi  -\frac{1}{4}R\phi)$ is invariant under $SU(2)'$, so we may want to use the same background charge changing deformations that we used in \eqref{inter}. As we have already mentioned, this is problematic because then the K\"ahler potential deformation $\partial \phi \bar{\partial}\phi e^{-\phi}$ will not be conformal invariant. Without a good justification except it works, we will use the other deformations $(-1+Q^{-2}){\partial}\phi \bar{\partial} \phi + (-1+Q^2)\partial Y \bar{\partial} Y $ that will make the K\"ahler potential Liouville interaction conformal invariant and make the central charge correct. As we discussed above, at the free level, these are physically the same deformations, so we may be able to declare that we simply chose the correct one after the $SU(2)'$ rotation.

Since $(-1+Q^{-2}) \partial \phi \bar{\partial} \phi$ is $SU(2)'$ invariant, we will focus on the action of $SU(2)'$ on $(-1+Q^2) \partial Y \bar{\partial}Y$.  
Now the term $\partial Y \bar{\partial}  Y = -(J_3-J_{3}')(\bar{J}_3-\bar{J}_3')$ is not $SU(2)'$ invariant so we should see the direct effect of the rotation here. After the rotation, we end up with
\begin{align}
(1-Q^2)\int d^2x (J_3-J_{1}')(\bar{J}_3-\bar{J}_{1}' ) \cr
= (1-Q^2) \int d^2x (\partial Y -\partial H - e^{iH+iY} -e^{-iH-iY})(\bar{\partial} Y - \bar{\partial} H - e^{iH+iY} - e^{-iH -iY} ) \ .
\end{align}
Note that this is a current-current deformation so that it is (exactly) marginal. By construction, we also have the Liouville K\"ahler potential deformation (without the superpotential deformation).

We see that the shift of $Y$ broken by the Liouville superpotential is not broken by the dual Liouville K\"ahler potential, but instead the shift of $H+Y$ (i.e. combination of the shift of $Y$ and the fermion number) is now broken by the rotated background charge changing deformations. As in the $\mathcal{N}=4$ case with $Q=1$, the name of the symmetry broken in the original frame and the dual-frame is different, but the physics must be the same. In particular, if we restrict ourselves to the charge-neutral correlation functions in which the shift of $Y$ is not broken, we may  readily compute the correlation functions as in the originally proposed duality and we should obtain the same result.

It is not obvious if the modified duality considered here are of practical use.  We, however, would like to point out that there have been several related studies since 2004. One interesting study is a (re)discovery of the relation between  the $SL(2,\mathbb{R})$ Wess-Zumino-Witten model and the bosonic Liouville theory \cite{Stoyanovsky:2000pg}\cite{Ribault:2005wp}\cite{Hikida:2007tq}. This relation may explain why there are poles associated with the screening by $e^{Q\phi}$ and $e^{-Q\phi}$ in the $\mathcal{N}=2$ supersymmetric Liouville theory by relating the correlation functions to the ones in the bosonic Liouville theory. Of course, this leads back to the original question of what we really mean by the duality in the (bosonic) Liouville theory. It would be fantastic if we could understand the physical meaning of it as Eguchi-san did in terms of the underlying geometric structure of the $A_1$ singularity for the $\mathcal{N}=2$ supersymmetric Liouville duality at $Q=1$.

\section{Conclusion}
I would like to share one of my personal recollections with Eguchi-san. In 2011, Eguchi-san was on sabbatical and visiting Caltech, where, at the time, I was working as a research assistant professor. One day, he asked me to accompany him to visit Griffith Observatory in L.A, which is about a one-hour drive from Pasadena. I thought he had better visit there with his family but he insisted, so I accompanied him.

The observatory has beautiful scenery, but what impressed us more was that in front of the observatory so many amateur astronomers got together, showing off their personal telescopes to look at the sky. I'm not sure if it was a special occasion, but it was spectacular. I've never seen such many amateur astronomers in one place. Eguchi-san had never failed to mention this amazing gathering whenever I met him after this visit.

I was still wondering what was so special about Griffith Observatory to Eguchi-san? Having Korean BBQ after our visit to the observatory, Eguchi-san told me the story. He had a fond memory of this Observatory because he once visited there with Nambu-san on their journey when he was a postdoc at Chicago. I realized that this is why he had to visit Griffith Observatory with me, but not with his family.

I was very honored to be a student of Prof. Tohru Eguchi.

\section*{Acknowledgement}
This article is based on my talk ``My unfinished work with Eguchi-san" at the workshop ``particle physics and mathematical physics -- forty years after Eguchi-Hanson solution". I thank the participants, in particular K.~Hori and Y.~Sugawara for valuable comments. I also thank S.~Murthy for the correspondence and sending me his fond memory with Eguchi-san. 

%%%%%%%%%%%%%%%%%%%%%%%%%%%%%%%%%%%%%%%%%%%%%%%%%%%%%%%%%%%%%%%%%%%%%%%%%%%%%%%%%%

\end{document}